# The International Committee for Future Accelerators (ICFA): History and the Future‡

Pushpalatha C. Bhat* and Roy Rubinstein**

Fermi National Accelerator Laboratory[1], Batavia, IL 60510, USA

*ICFA Secretary 2016-Present
**ICFA Secretary 1993-2016

## 1. Introduction

The International Committee for Future Accelerators (ICFA) has been in existence now for four decades. It plays an important role in facilitating discussions and collaboration within the world particle physics community on the status and future of very large particle accelerators as well as the particle physics and related fields associated with them. This article traces the origins of ICFA, its structure and mandate, its activities and accomplishments, and its anticipated activities related to the future directions of the field of particle physics.

The ICFA website (http://icfa.fnal.gov/) contains much additional information, including ICFA membership, ICFA Statements, and documents leading to the current design of the International Linear Collider.

## 2. Origins

The origins of ICFA go back to the late 1960s. A series of East-West international meetings was held during 1967-1976 to review future perspectives in particle physics. By the end of that period, there was a belief that the next large accelerator, after the Fermilab and CERN 400 GeV proton synchrotrons, would of necessity, because of its

---

complexity and cost, be an international machine. A key meeting in ICFA's formation took place in New Orleans in 1975. Some fifty particle physics world leaders passed a resolution recommending the formation of a group to study the scientific, technical and organizational problems connected with world-wide collaboration in the construction of a very large accelerator. This recommendation led IUPAP's Commission 11 (C 11, Particles and Fields) to create ICFA in 1976, with its formal establishment in 1977.

Included in the original C 11 terms of reference for ICFA were:

"To organize workshops for the study of problems related to an international super high energy accelerator complex (VBA: Very Big Accelerator) and to elaborate the framework of its construction and of its use" (the VBA was envisioned as a ~ 20 TeV proton accelerator which could allow colliding beams), and

"To organize meetings for the exchange of information on future plans of regional facilities and for the formulation of advice on joint studies and uses".

## 3.  ICFA Membership and Meetings

ICFA membership is approximately representative of particle physics activity in the different regions of the world. Since 1995, this has been (member numbers in parentheses): CERN member states (3), USA (3), Japan (2), Russia (2), Canada (1), China (1), Other Countries (3); the Chair of C 11 is an ex-officio ICFA member. ICFA members are nominated by designated authorities in their countries or regions, followed by C11 confirmation. Member terms are 3 years, and can be renewed. The ICFA Chair (a three year term) is chosen by the ICFA membership.

ICFA meetings are typically held twice a year, with invitations to the longer of the two meetings going also to the directors of all of the world's major particle physics labs; this allows a much more extensive discussion of the current status and future of particle physics.

ICFA meetings have led to many ICFA Statements (see the ICFA website) and documents such as "Beacons of Discovery" (2011) intended to convey aspects of particle physics to broader audiences.

There is close cooperation between ICFA and FALC (a group of representatives of international particle physics funding agencies); the chairs of each organization attend the other's meetings.

## 4. ICFA Guidelines for Particle Physics Facilities

In 1980, ICFA produced Guidelines for the utilization of major regional facilities for particle physics research; these were reaffirmed in 1993, with an update in 2011. Two of the current Guidelines are:

"The criteria used in selecting experiments and determining their priority are (a) scientific merit, (b) technical feasibility, (c) capability of the experimental group, (d) availability of the resources required"

"Operating laboratories should not require experimental groups to contribute to the running costs of the accelerators or colliding beam machines nor to the costs of their associated experimental areas. However, in particular for a large global facility, allocation of operating costs should be agreed by the project partners before project approval, while still allowing open access for experimental groups".

## 5. ICFA Panels

There are accelerator and particle physics topics of a technical nature where international discussion and collaboration is needed, and where the required expertise is beyond that of the individual ICFA members. Because of this, ICFA has set up several Panels, each with ~16 experts from around the world, on specific technical areas. The topics of the current Panels are: Instrumentation; beam dynamics; advanced and novel

P.C. Bhat, R. Rubinsteinaccelerators; interregional connectivity; particle physics data preservation; the linear collider board; sustainable accelerators and colliders.

A panel on accelerator-based neutrino facilities completed its mandate and produced a final report in 2017. (More details in Section 11.)

Each Panel organizes its own program, which can include workshops, newsletters, schools, etc., and each Panel regularly reports at ICFA meetings.

6.  **ICFA Seminars**

ICFA Seminars, with the title "Future Perspectives in High-Energy Physics", have been held every 3 years since 1984. They generally run for three and a half days, usually take place at a major particle physics laboratory, and have an invitation-only world-wide attendance of ~150 to 200; government science officials and media representatives are also invited. Review talks are given at the Seminars on the status and prospects of accelerators, particle physics, and related fields around the world.

7.  **ICFA and the SSC**

At the August 1983 ICFA meeting, there was much concern expressed about the newly announced US Superconducting Super Collider (SSC), a proton-proton collider with a center of mass energy of 40 TeV (in a 87 km circular tunnel) which, as a national project, was not compatible with ICFA's goal, noted in Section 2 above, that such a machine be international. This concern led to the organization of the first ICFA Seminar in May 1984, and the SSC was the central topic of discussion at the Seminar, leading to the ICFA Chair's summary including the statement:  "ICFA views its major role as facilitating the construction of high-energy accelerators and not as arbitrating among various national or regional options".

In 1985, ICFA redefined its aims to be:

"To promote international collaboration in all phases of the construction and exploitation of very high energy accelerators"

"To organize regularly world-inclusive meetings for the exchange of information on future plans for regional facilities and for the formulation of advice on joint studies and uses"

"To organize workshops for the study of problems related to super high-energy accelerator complexes and their international exploitation and to foster research and development of necessary technology".

## 8. Termination of the SSC and Approval of the LHC

In October 1993 the SSC, under construction in the US, was cancelled. Within days, many in the world particle physics community urged ICFA to start discussions on the field's situation following this cancellation. This quickly led to an ICFA statement noting the world-wide interest in participation in the LHC, and urging the CERN Council to find appropriate mechanisms to bring non-member states into the project. Following the initial LHC approval by the CERN Council, in January 1995 ICFA stated it "warmly welcomes the approval of the ….LHC" and in October 1996 noted that "the LHC…. is essential for the progress of the field" and that "the LHC is becoming a true world facility". In January 1997, ICFA noted "with great satisfaction" that the LHC was now approved for construction in a single stage.

## 9. Global Design of a Linear Collider, the ILC

By the early 1990s, a consensus was emerging among the particle physics communities of the world that a linear $e^+e^-$ collider with center of mass energy in the 100s of GeV to a TeV was the next major accelerator needed for the field following the Tevatron, SSC and LHC hadron colliders. The issue was raised often at ICFA meetings in the 1990s, and several ICFA Statements from 1993 onwards reiterated this consensus. Since no nation was ready to take the lead on the R&D and design of this machine, ICFA itself from ~2000 onwards organized linear collider activities for the world particle physics community, and has continued to do so to the present time.

In 2002, ICFA formed the International Linear Collider Steering Committee (ILCSC) to facilitate the realization of the International Linear Collider (ILC) as a global collaborative effort.

In 2003 a Parameters Committee set up by the ILCSC produced a recommended set of performance parameters for the ILC, which was subsequently updated in 2006; these parameters were used by the GDE (see below) as ILC design criteria.

By 2004, there were two major technologies being pursued in the world for the main linacs of a linear collider with energy range up to 1 TeV: room temperature X-band (11.4 GHz) and superconducting L-band (1.3 GHz), and a choice between the two needed to be made. ILCSC set up the International Technology Recommendation Panel in 2003, chaired by Barry Barish, to recommend which technology to pursue. The Panel's recommendation in its 2004 report was to go forward with the superconducting option; this was immediately accepted by ICFA, and the world particle physics community quickly united behind this technology choice.

The ILCSC set up the Global Design Effort (GDE) in 2005, with Barry Barish as Director, to produce a technical design for the ILC. The Technical Design Report was completed in 2012 and published in 2013; following successful technical and cost reviews of the design, the GDE and ILCSC both went out of existence in 2013.

In 2013, ICFA established the Linear Collider Board (LCB) to succeed ILCSC; in addition to overseeing the work on the ILC, the LCB also oversees the CLIC project (a linear collider with potential for higher energies than ILC, but with several years more R&D still needed) and the CLIC and ILC detectors. Under the LCB is the Linear Collider Collaboration (LCC), with Lyn Evans as Director, to coordinate and direct the global effort towards realizing a linear collider.

P.C. Bhat, R. Rubinstein

## 10. Particle Physics Driving the Future

Historically, progress in particle physics at accelerators has come from experiments performed at successively higher collision energies and with high intensity beams. Because of their energy reach, hadron machines have always been discovery machines. Examples are the discoveries of W and Z bosons at the CERN $S p\bar{p}S$ collider, the discoveries of bottom and top quarks and the tau neutrino at the Fermilab Tevatron, and the discovery of the Higgs boson at the CERN LHC. Lepton colliders, on the other hand, provide exquisitely precise measurements; e.g. LEP enabled high precision measurements of many Standard Model quantities.

The discovery of the Higgs boson in 2012 by the CMS and ATLAS experiments at CERN's LHC completes the Standard Model (SM) of particle physics. All matter and force-carrier particles expected in the SM have now been discovered; a majority of them in experiments at successively more powerful particle accelerators and colliders, and the predictions of the SM have been confirmed with remarkable precision in innumerable measurements.

While the discovery of the Higgs boson was a crowning achievement for the Standard Model, it could also be the harbinger of new physics beyond the Standard Model (BSM). Despite its spectacular success, the SM does not provide answers to several fundamental questions such as the origin of neutrino mass and oscillations, the existence of three generations of quarks and leptons, the dominance of matter over antimatter in the universe, the nature of dark matter, etc. Therefore, the search for BSM physics must continue! The Higgs discovery has reignited strong interest in the world-wide particle physics community to develop and build next generation collider facilities beyond the LHC to study the properties of the Higgs boson with great precision as well as to access BSM physics.

The LHC, which started stable operations in 2010 has a lot more physics to offer and possibly more discoveries in the coming decades. By the end of 2023, the LHC will have delivered ~300 fb$^{-1}$ each to the CMS and ATLAS experiments. Then the LHC and the detectors will undergo major upgrades into High Luminosity LHC (HL-LHC), increasing the collider luminosity by a factor of five over the initial LHC design, with the goal of delivering ~3000 fb$^{-1}$ by 2035.

P.C. Bhat, R. Rubinstein

Our present limited understanding has also highlighted the need to investigate the neutrino sector in detail (since some observed neutrino properties are not predicted by the SM) and also rare processes with high precision using very high intensity particle beams. The discovery of neutrino flavor oscillations by the Super Kamiokande experiment in 1998 was a seminal breakthrough in neutrino physics. Since then, steady progress is being made in the studies of neutrino interactions and oscillation physics. Planning of long-term experimental programs is underway to explore and understand neutrino masses, mixing and CP violation in the neutrino sector.

ICFA is heavily involved in world-wide discussions on future accelerator facilities needed to answer the particle physics questions outlined above; these discussions occur at ICFA meetings, at ICFA Seminars, and through ICFA's Panels.

At the energy frontier, two categories of future colliders are currently of great interest: (1) a "Higgs Factory", a lepton collider, either linear or circular, with a center of mass energy of 250 GeV and above, to perform precision studies of the Higgs boson properties including its couplings to other SM particles; and (2) a ~100 TeV scale proton-proton collider to explore the "next energy frontier" to elucidate electroweak symmetry breaking (EWSB) and to search for BSM physics and new particles up to ~50 TeV mass scale. Multi-TeV lepton colliders ($e^+e^-$ or $\mu^+\mu^-$) are another category of future colliders that would enable searches and studies of new physics signals in the TeV range, in a way complementary to hadron colliders.

There are several excellent candidates for such future facilities currently under various stages of consideration and planning in different regions of the world. ICFA has been following the studies and facilitating the planning of such major international facilities. These include the International Linear Collider (ILC), discussed earlier and currently being considered for construction in Japan; the Compact Linear Collider (CLIC); the Future Circular Collider (FCC) being investigated at CERN (FCC has options for $e^+e^-$, pp and ep collisions); the Circular Electron Positron Collider (CEPC) and the Super Proton-Proton Collider (SPPC) in China; and the Long Baseline Neutrino Facility (LBNF) in the US and Hyper-Kamiokande in Japan.

## 11. Neutrino Facilities

Measurements of parameters that govern neutrino oscillations significantly impact our understanding of particle physics, astrophysics and cosmology. Therefore, ambitious and intense efforts are underway worldwide charting out the required experimental program. Many short-baseline (SBN program at Fermilab) and long baseline (NOvA at Fermilab and T2K in Japan) oscillation experiments are ongoing and international large scale projects, i.e., the Long Baseline Neutrino Facility (LBNF) with the Deep Underground Neutrino Experiment (DUNE) in the United States and Hyper Kamiokande (Hyper-K) in Japan, are in planning and preparation. Since accelerator-based neutrino experiments are an essential component of the program, ICFA created a Neutrino Panel in 2013 with the mandate to "promote international cooperation in the development of the accelerator-based neutrino oscillation program and to promote international collaboration in the development of a neutrino factory as a future intense source of neutrinos for particle physics experiments."

The ICFA Neutrino Panel completed its study in 2017 and produced a "Roadmap" document that can be found on the ICFA web page (http://icfa.fnal.gov/panels/neutrino_panel/). The Panel, in its initial report, outlined a program by which the discovery potential of the accelerator-based neutrino program could be optimized. The Panel stressed that greater international cooperation will be the key to the success of the program. The establishment of the SBN and LBNF programs at Fermilab and the CERN Neutrino Platform as international facilities represents substantial progress in the necessary internationalization of the program. The Panel also noted the recent developments in the international Hyper-K program. Together, the complementary long-baseline experiments DUNE and Hyper-K, the SBN experiments and the CERN Neutrino Platform will provide the basis for a robust discovery program well into the 2030s. The results should provide us with deep insights into the mixing among the three neutrino flavors, the ordering of the neutrino mass eigenstates, possible CP violation, and whether the measurements can be accommodated within the elegant framework of a three-neutrino mixing model, or yield new discoveries.

## 12. Future Lepton Colliders

Since the discovery of the Higgs boson, the global HEP community has become convinced that a lepton collider to study the Higgs in great detail and with exquisite precision is necessary. In contrast to hadrons, leptons are elementary point-like particles, and lepton colliders provide distinct advantages of well-determined initial state properties and simple final states in a cleaner environment for study of particle properties and new phenomena.

A lepton collider Higgs factory could be either a circular collider ($e^+e^-$ or $\mu^+\mu^-$) or a linear collider ($e^+e^-$). A circular collider in a large enough ring, such as the FCC-ee or CEPC, operating at energies in the range of 240-400 GeV would use well established technology and can provide a high luminosity Higgs Factory for studies at the Higgs production threshold to ttbar production threshold. However, the advantage of a linear $e^+e^-$ collider is that it could be upgradeable to much higher energies, to 1 TeV (in the case of ILC). Measurements at these higher collision energies significantly improve the precision of measurements of Higgs couplings to SM particles and especially of the Higgs self-coupling. The LHC and HL-LHC data to be accumulated over the next two decades will continue to improve our knowledge of the Higgs boson. But it is well recognized that the precision to which we need to know the coupling strengths of the Higgs and to access BSM physics will require colliders beyond the LHC.

12.1 The International Linear Collider

The ILC and ICFA's engagement in enabling its global design have been discussed in Section 9. The discovery of the Higgs boson significantly strengthened the physics case for the ILC, and the Japanese particle physics community has expressed strong interest to host the ILC. Over the past few years, substantial technical progress has been made. Fermilab's superconducting RF (SRF) R&D program has exceeded the ILC accelerating gradient specification of 31.5 MV/m. In addition, the European XFEL at DESY is recognized as a successful large-scale prototype of an SRF accelerator for the ILC.

While the technical challenges seem tractable and can be overcome, funding for ILC construction as an international project still needs to be determined. In the past two years, the LCC, together with the Japanese particle physics community, has extensively studied various staging scenarios of the ILC machine and their physics potential and implications, and the LCB has proposed that a 250 GeV ILC be constructed in Japan, as soon as feasible. With its 40% initial cost reduction, and its future extendibility to higher energies, the 250 GeV ILC is now viable as the next big international facility for particle physics research. ICFA endorsed the LCB recommendation in 2017.

P.C. Bhat, R. Rubinstein

An updated staging baseline document for CLIC, for construction and operation in three stages, the first stage being a 380 GeV Higgs factory has now been released. An updated cost estimate for a low energy Higgs factory revealed that the costs are very similar to the costs for an ILC with similar energies.

12.2 Multi-TeV Lepton Colliders

The candidate for a potential multi-TeV (up to 3 TeV) $e^+e^-$ collider would be CLIC, which is based on a novel two-beam acceleration using 12 GHz X-band normal conducting (NC) RF structures. The technical feasibility of the two-beam acceleration has been demonstrated recently in a small scale CLIC Test Facility (CTF3) at CERN, where average accelerating gradients in excess of 100 MV/m were achieved with acceptable RF cavity breakdown rates. The biggest issues, however, are the enormous site power requirement (~600 MW) and the anticipated cost.

While an $e^+e^-$ collider such as CLIC can provide an energy reach up to 3 TeV, a muon collider, if feasible, can reach well above 3 TeV to energies possibly up to 10 TeV. The synchrotron radiation is significantly smaller (by ~$10^9$) and the production cross section for s-channel Higgs is ~40,000 times larger than in case of $e^+e^-$. It will be relatively compact and is expected to be relatively much more cost-effective and affordable. Unfortunately, at present, the performance of the muon collider can be assured only at the level of two to three orders of magnitude below the design luminosity goal of $2\cdot10^{34}$ cm$^{-2}$s$^{-1}$ and the performance feasibility requires convincing demonstration of the 6-D ionization cooling of muons. The MICE experiment at RAL provided the first experimental evidence of muon cooling in 2018.

There are novel ideas being explored, however, to realize the muon beam quality required for a collider. Multi-TeV lepton colliders, including a muon collider, will be of great interest if the LHC finds any resonances in the TeV mass range. Studies have shown that the technology involved in realizing a muon collider could potentially also lead to the development of a very high intensity neutrino factory.

**13. Future Hadron Colliders**

Following the European Particle Physics Strategy recommendation in 2013, CERN has been leading a study of a Future Circular Collider (FCC) project at CERN, in a 100 km tunnel, housing a 100 TeV proton-proton collider. The FCC project also proposes to use the tunnel first to house an $e^+e^-$ collider with collision energy ~240-350 GeV to study the Higgs boson, followed by the installation and commissioning of the 100 TeV pp collider, and eventually an option of an e-p collider. More recently, CERN has integrated the study of a High Energy LHC (HE-LHC) with the FCC project. The idea is that the replacement of the magnets in the current LHC tunnel with 16-20 T magnets can provide a machine, with a design collision energy of 27-33 TeV. The FCC project is expected to produce a Conceptual Design Report (CDR) in 2018 in preparation for the next round of European Particle Physics Strategy exercise in 2019.

As mentioned in Section 10, China is also planning and carrying out design studies for a Super Proton-Proton Collider (SPPC) possibly as the second stage project in the CEPC tunnel. A preliminary CDR on the accelerators was released by the CEPC-SPPC Study Group in March 2015. A CDR is expected to be released in 2018.

For hadron colliders beyond the LHC such as FCC-hh (which may collide protons or heavier ions), HE-LHC and SPPC, to be feasible, several technological challenges must be overcome. To operate at the proposed collision energies would require development of ~16-20T superconducting (SC) magnets; vigorous R&D on $Nb_3Sn$ superconductor technology and of advanced hybrid magnets which use Nb-Ti, $Nb_3Sn$ and High Temperature Superconductors (HTS) in the inner layer are being carried out. There are other critical challenges. The biggest hurdle for embarking on construction of such large machines with ~100 km circumference is their cost. One of the primary goals of the long-term high field magnet (16-20 T) R&D program, therefore, has to be significant reduction in cost per TeV.

### 14. Far-Future Colliders

The next generation of very high energy colliders, beyond the ones discussed above, will require significant breakthroughs in acceleration techniques and on other fronts. There has been much hope and anticipation that plasma-based accelerators, either beam-driven or laser-driven wake-fields that provide accelerating gradients of the order of tens of GV/m, will enable the construction of more compact and affordable colliders. There are proof-of-principle experiments of the wakefield acceleration concept but several formidable challenges such as beam stability and control, emittance and brightness preservation, staging of the accelerator, etc. have to be overcome to build such an advanced and compact accelerator for particle physics.

ICFA has created a panel on Advanced and Novel Accelerators (ANA) to monitor progress in this area and hold studies and workshops to facilitate communication and exchange of ideas within the community, and promote international collaboration. With the availability of lasers with multi-terawatt power and a few tens of femtosecond–long pulses, the panel also has the mandate to monitor opportunities for the application of advanced and novel accelerator technology beyond particle and nuclear physics such as in medical physics, nondestructive evaluation, and security.

The ICFA ANA panel recently held a workshop focused on exploring novel acceleration techniques and available technology for an electron/positron ($e^-e^+$) collider or an electron/proton (e-p) collider, both at the energy frontier. The workshop was organized as a first step towards the development of an international ANA scientific roadmap for such an advanced linear collider, with the delivery of a technical design report by 2035. The first step towards this goal includes taking stock of the scientific landscape, outlining global priorities for scientific progress, identifying facilities necessary for this progress, identifying existing and local roadmaps, and developing strategies to encourage the development of a genuinely international R&D roadmap. ICFA very much supports the panel's efforts to bring together the advanced acceleration research community to develop the R&D roadmaps for each of the candidate techniques with common milestones and eventually developing criteria for selection among the techniques.

## 15. Summary

In the twenty-first century, particle physics has emerged as a truly global enterprise for discovery in fundamental science. ICFA has been playing an important role as a forum for discussions transcending national or regional boundaries on the future of high energy accelerators as global research facilities and their associated particle physics, detectors, and technology. It is probably true that, to paraphrase an old expression, if ICFA didn't exist, something very similar would have to be invented. As also appeared to be true in the 1970s, projects under consideration in the field are becoming so large and costly that no single country or a group of countries can carry them out in isolation.

ICFA facilitated the international collaboration in the LHC in the aftermath of the cancellation of the SSC, enabled the global design effort of the ILC, unified the ILC and CLIC communities to carry out coordinated studies and R&D, and has recently championed the case for a Higgs Factory ILC in Japan. ICFA has been very relevant for international cooperation and is anticipated to play a critical role in promoting international collaboration in all phases of construction and exploitation of future global accelerator facilities for particle physics.

## 16. Acknowledgments

The authors thank the editors of the volume for the invitation to write this paper, and also thank Joachim Mnich and Geoff Taylor, the past and current chairs of ICFA, for useful comments during the preparation of this paper.

**Pushpalatha Bhat** is a Senior Scientist at Fermilab. Her research career has spanned physics from keV energies to the energy frontier. She has held many leadership roles at Fermilab including the successful luminosity upgrades of the Tevatron accelerator complex. She pioneered the application of machine learning methods in high energy physics, and made significant contributions to the discovery of the top quark at the Tevatron and the Higgs boson at the LHC. She is a Fellow of the American Physical Society and the American Association for the Advancement of Science. She currently serves on the APS Board of Directors and the APS Council, and as Secretary of ICFA.

**Roy Rubinstein** has carried out particle physics experiments at the Birmingham (UK) proton synchrotron, the Brookhaven AGS, and the Main Ring and the Tevatron Collider at Fermilab. Much of that work was measuring the hadron-hadron elastic scattering and total cross section over a wide range of energies. He was associated with Fermilab from contributing to the initial 1968 design report to his retirement as Assistant Director in 2016. For several decades, he was Fermilab's "Foreign Minister". He served as the Secretary of ICFA from 1993-2016.